\documentstyle[prl,aps,epsfig,multicol]{revtex}
\begin{document}
\title{Colloidal interactions in two dimensional nematics}

\author{M. Tasinkevych$^1$, N.M. Silvestre$^1$, P. Patr\'\i cio$^{1,2}$, and M.M. Telo da Gama$^1$} 
\address{$^1$Departamento de F{\'\i}sica da Faculdade de Ci{\^e}ncias and 
Centro de F{\'\i}sica Te\'orica e Computacional\\
Universidade de Lisboa, 
Avenida Professor Gama Pinto 2, P-1649-003 Lisboa Codex, Portugal\\
$^2$Instituto Superior de Engenharia de Lisboa\\
Rua Conselheiro Em\'\i dio Navarro 1, P-1949-014 Lisboa, Portugal
}

\date{September 2002}

\maketitle

\begin{abstract}

The interaction between two disks immersed in a 2D nematic is
investigated
(i) analitically using the tensor order parameter formalism for the
nematic configuration around isolated disks and
(ii) numerically using finite element methods with adaptive meshing
to minimize the corresponding Landau-de Gennes free energy.
For strong homeotropic anchoring, each disk generates a pair of defects
with one-half topological charge responsible for the 2D quadrupolar
interaction between the disks at large distances.
At short distance, the position of the defects may change, 
leading to unexpected complex interactions with the quadrupolar repulsive
interactions becoming attractive. This short range attraction in all
directions is still anisotropic.
As the distance between the disks decreases their preferred relative 
orientation with respect to the far-field nematic director changes from 
oblique to perpendicular.

\end{abstract}

\vspace{2mm}
PACS numbers: 77.84.Nh Liquids, emulsions, and suspensions; liquid crystals -
61.30.cz Theory and models of liquid crystal structure - 
61.30.Jf Defects in liquid crystals.

\begin{multicols}{2}

\section{Introduction}

In the last few years, there has been considerable theoretical and 
experimental interest in colloidal systems with
anisotropic host fluids \cite{Stark_review}.

Since the temperature changes the anisotropic properties
of the host fluid, the dispersed particles exhibit
a rich variety of collective behaviors, 
leading to unexpected phase diagrams \cite{Yamamoto}.
Depending on the details of the system,
chain-like \cite{Poulin_nature,Loudet_nature}, 
crystal \cite{Nazarenko} or cellular \cite{Anderson} structures
have been observed.

In the case of nematic hosts in three dimensions (3D), the colloidal
interactions result from the competition between the anchoring properties
of the spherical colloidal surfaces favoring e.g., 
radial (homeotropic) nematic orientation (anchoring)
and the bulk elasticity that favors a uniform nematic.
For sufficiently strong homeotropic anchoring $W$, 
the spherical colloids may induce topological defects or singularities in
the nematic director field.
If the size of the colloidal particle $a$ is large 
when compared to the nematic correlation length $\zeta$, 
a hyperbolic point defect appears at a certain distance from the particle
({\sl satellite dipolar} configuration 
\cite{Ruhwandl,Lubensky,Stark,Andrienko}).
For smaller particles, however, a ring disclination around
the equatorial plane of the sphere may have a lower energy 
({\sl saturn-ring quadrupolar} configuration \cite{Gu}). Finally, 
if $Wa$ is smaller than the typical nematic elastic constant $K$, 
the director field varies smoothly and there are no defects 
({\sl surface-ring quadrupolar} configuration \cite{Kuksenok}).
Depending on the symmetry of the distortion \cite{Lev},
 the dominant long range interactions are dipolar \cite{Poulin_prl} 
or quadrupolar \cite{Ramaswamy,Mondain-Monval}.
The short range interactions are more difficult to predict,
as nonlinear terms in the elastic free energy come into play.
Experimental evidence from a system where silicon oil
droplets were dispersed in a nematic host 
\cite{Loudet_prl} reveals that when the nematic configuration around each
droplet has dipolar symmetry,
the attractive long range interaction between aligned droplets 
becomes strongly repulsive at short distances.
A nematic topological defect appears between the particles and prevents
their collapse.
By contrast, when the long range distortion exhibits quadrupolar symmetry 
no such defects appear and the oil droplets coalesce.

2D colloidal systems were also investigated
recently and similar qualitative features were observed.
In these systems the interactions between the dispersed particles
are also mediated by the elastic distorsions of the anisotropic host.
However, for 2D nematic hosts and homeotropic anchoring conditions,
only quadrupolar configurations are stable \cite{Fukuda}.
Dipolar configurations may be found in 2D smectic C films \cite{Cluzeau},
a system that is similar to a 2D nematic,
but where the director does not exhibit mirror symmetry,
thus excluding configurations with half integer topological charges.
In this case, Pettey et al. \cite{Pettey} obtained 
analytic expressions for the free energy
of a system of multiple disk-defect pairs, 
valid in the linear regime where the particles are sufficiently far apart.
We investigated the short range interactions in this system 
by using a numerical method capable
of describing the nematic orientation profiles 
induced by a small number of colloids with arbitrary shape in 2D 
\cite{Patricio}.
Numerical investigations of this type are faced with a problem due
to the very different length scales characterizing the colloids and the 
topological defects. We have solved it \cite{Patricio}, by using
finite elements with adaptive meshing to minimize the Frank free 
energy and obtained very accurate interaction energies between disks
at arbitrary separations.
We have found a strong repulsion at short range followed by the expected
long range dipolar interaction with a pronounced minimum at 
an intermediate distance. The equilibrium director field
configuration exhibits a topological defect at the mid-point between the
disks.

In this article, we use similar numerical methods \cite{Patricio}
to minimize the Landau-de Gennes free energy of two circular disks
in a 2D nematic.
The great advantage of this model is that no special treatment of the
topological defects is required, by contrast with the Frank elastic free
energy description, where the defects appear as singularities
of the director field ${\bf n}$.
In the next section, we describe the nematic free energy based on the
tensor order parameter formalism.
Then, we generalize the analytical results of Pettey et al. \cite{Pettey}
to nematic configurations with half integer topological defects and obtain
analytic expressions for the free energy of the quadrupolar
nematic distortion for a single disk and the long range
quadrupolar interaction between disks.
Finally, we present the results of extensive numerical calculations for
the interaction between disks at arbitrary separations.

\section{The Landau-de Gennes free energy}

In the tensor order parameter formalism the 2D nematic order is
represented by the (tilt) angle of the director field ${\bf
n}=(\cos\theta,\sin\theta)$, that describes the mean orientation of the
molecules, and the orientational order parameter $Q$,
that measures the degree of alignment of the molecules with respect to the
mean orientation.
In the isotropic phase, where the molecules point in every direction, 
$Q=0$.
By contrast, in the nematic phase the molecules are preferentially aligned
and the orientational order parameter takes a well defined equilibrium
value $Q=Q_{eq}$ that depends on the temperature.

We define a 2D symmetric traceless tensor field, from the
tilt angle
$\theta({\bf r})$ and the orientational order parameter $Q({\bf r})$, 
\begin{equation}
Q_{ij}({\bf r})=Q({\bf r})\left(n_i({\bf r})n_j({\bf r})-\delta_{ij}/2\right) 
\label{eq_tensor_definition}
\end{equation}
that is invariant under the nematic symmetry operations 
${\bf n\rightarrow  -n}$ and ${\bf r \rightarrow -r}$.

The Landau-de Gennes free energy may be understood as
a Taylor expansion in terms of the tensor order parameter $Q_{ij}$ (the bulk 
free energy)
and its derivatives $\partial_kQ_{ij}$ (the elastic free energy)
\cite{deGennes}:
\begin{equation}
F=\int_\Omega \Big(-\frac{A}{2}Q_{ij}^2+\frac{C}{4}(Q_{ij}^2)^2+ 
\frac{L}{2}(\partial_kQ_{ij})^2\Big)d^2{\bf r} 
\label{eq_tensor_energy}
\end{equation}
where the summation over repeated indices is implied.
The integration is over the 2D space occupied by the nematic $\Omega$. 
Only terms with the nematic symmetry are allowed in the free
energy expansion. In addition, stability requires that the total free
energy is bounded from below. In the nematic $A$ and $C$ are positive
constants, defining the equilibrium value of the orientational order
parameter 
$Q_{eq}=\sqrt{2A/C}$.
For simplicity, we adopted the one-constant approximation for the elastic
free energy.
In the general 2D case two elastic terms --
corresponding to different invariant derivatives
of the tensor order parameter -- have to be included in the expansion. 
The elastic constant $L$ is related to the Frank elastic constant through 
$K=4LA/C$.
The total free energy can be written as a functional of the tilt angle
$\theta$ and the orientational order parameter $Q$:
\begin{equation}
F=\int_\Omega \Big(-{A \over 4}Q^2+{C \over 16}Q^4+{L \over 4}(\nabla
Q) ^2+LQ^2(\nabla\theta)^2 \Big)d^2{\bf r}.
\label{eq_energy}
\end{equation}

Let $a$ be a typical length of the system, e.g. the radius of the 
colloidal particles.
If $L/a^2\ll A$, the energy associated with the variation of the tilt 
angle $\theta$ is much smaller than the energy associated with the
orientational order parameter.
In this case we may set $Q=Q_{eq}$ everywhere in the nematic and the free
energy simplifies to
\begin{equation}
F_{el}=\frac{K}{2} \int_\Omega (\nabla\theta)^2 d^2{\bf r}
\label{eq_frank_energy}
\end{equation}
which is the one-constant elastic Frank free energy.

For strong distortions, however, topological defects may appear. In this
case the director field becomes singular and within a small region around
the defect the orientational order parameter vanishes and the contribution
of the first three terms in Eq.(\ref{eq_energy}) has to be taken into
account.
This energy -- the {\sl core energy} of the defect --
arises from a small region of size $\xi$, and may be estimated
by a boundary layer analysis \cite{Chaikin}.

Let us consider a topological defect of charge $q$ placed at the center of
the 2D space.
In polar coordinates ${\bf r}=(r,\phi)$, the defect may be described by
$\theta(\phi)=q\phi$, asymptotically valid in the core region $r<\xi$,
far from any other boundaries.
The orientational order parameter, constant outside the core region,
decreases rapidly to $Q(r=0)=0$ inside the core in order to regularize
the nematic free energy.
A simple description of this behavior of $Q$ is given by the ansatz
\begin{equation}
Q(r)=Q_{eq}(1-e^{-\frac{r}{\xi}}) \;\; .
\label{eq_q_approximation}
\end{equation}
By substituting this expression into the free energy (\ref{eq_energy})
we obtain the simple result
\begin{equation}
F(\xi)=F_c(\xi)+F_{el}(\xi)=f\frac{A^2}{C} \xi^2+q^2\pi K \ln\frac{R_\infty}{\xi}
\label{eq_approximated_energy}
\end{equation}
where $f\approx 0.97$ and $R_\infty$ is the size of the system.
Differentiation with respect to $\xi$ yields $\xi =|q|\zeta$,
where $\zeta=\sqrt{(2\pi/f)(L/A)}$ is the nematic correlation length.
If $\zeta$ is much smaller than the other characteristic lengths of the system, 
we can use the Frank elastic free energy (\ref{eq_frank_energy}) to
obtain the director configuration in systems with topological defects.
The Frank elastic free energy, however, is defined outside the
core region only. The total free energy can be computed by additing to it
the core energy of each defect 
\begin{equation}
F_c=q^2\frac{\pi K}{2} \;\;.
\label{eq_core_energy}
\end{equation}

\section{Isolated disk}

Fukuda and Yokoyama \cite{Fukuda} 
studied the distortion of a 2D nematic around an isolated disk, with
homeotropic anchoring, by minimizing
the Landau-de Gennes free energy (\ref{eq_tensor_energy})
using numerical methods with adaptive grids.
They have shown that a director configuration with a dipolar defect
(topological charge $q=-1$) is unstable,
and changes into a configuration with a pair of defects with topological
charge $q=-1/2$ that are located symmetrically around the particle.

The energy of this quadrupolar configuration may be calculated
analitically and compared with the energy of the dipolar case, obtained by
Pettey et. al. \cite{Pettey} for 2D smectic-C films.

\begin{figure}[ht]
\par\columnwidth=20.5pc
\hsize\columnwidth\global\linewidth\columnwidth   
\displaywidth\columnwidth
\centerline{\epsfxsize=240pt\epsfbox{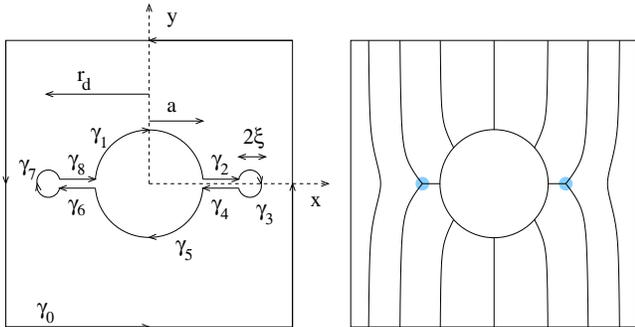}}
\caption
{Single disk. Left: notation and reference frame
used in this work. Right: quadrupolar configuration with two half integer
defects (grey). 
The lines represent the orientation of the nematic director.}
\label{fig_f1}
\end{figure}

Consider the disk at the center of the reference frame ${\bf
r}=(x,y)$,
or  ${\bf r}=(r,\phi)$ in polar coordinates, where $\phi$ is defined
counterclockwise, starting from the $x$-axis, in the range $-\pi<\phi<\pi$
(see Fig.\ref{fig_f1}).
At the outer boundary, for large $r$, the nematic is uniform with
$\theta=\pi/2$. Homeotropic boundary conditions apply at the disk's
boundary, i.e., $\theta(\phi)=\phi$ on the upper contour and 
$\theta(\phi)=\pi+\phi$ on the lower one.
Minimization of the Frank elastic free energy
(\ref{eq_frank_energy}) yields Laplace's equation $\Delta\theta=0$,
everywhere (in the nematic) except within the defects.

An image solution with two half integer topological charges
located symmetrically at a distance $r_d$ from the center of the disk,
satisfying the homeotropic boundary conditions at the disk's contour
and a uniform director at infinity, is given by
\begin{equation}
\theta=\frac{\pi}{2}-2\psi+\frac{1}{2}\psi_{r_d}+\frac{1}{2}\psi_{-r_d}
+\frac{1}{2}\psi_{\tilde{r}_d}+\frac{1}{2}\psi_{-\tilde{r}_d} 
\label{eq_quadrupolar_solution}
\end{equation}
where $\tilde{r}_d=a^2/r_d$ is the position of a virtual half integer
topological defect inside the disk. The field
\begin{equation}
\psi_{x_i}(x,y)=\arctan\frac{x-x_i}{y}
\label{eq_defect_solution}
\end{equation}
describes a topological defect of charge $q=1$ located at 
${\bf r}_i=(x_i,0)$ and satisfies Laplace's equation everywhere except at 
${\bf r}_i$.

The elastic Frank free energy can be calculated using Green's
Theorem:
\begin{equation}
 \int_\Omega (\nabla\theta)^2 d^2{\bf r}=
\oint_{\partial\Omega}\theta\frac{\partial\theta}{\partial {\bf n}}d{\bf l}
-\int_{\Omega}\theta\Delta\theta d^2{\bf r} \;\;,
\label{eq_green_th}
\end{equation}
where $\partial\Omega$ is the boundary of 
the 2D nematic domain, $d{\bf l}=(dx,dy)$ is an oriented contour element,
and $\partial\theta/\partial {\bf n}=(-\partial\theta/\partial y,\partial\theta/\partial x)$. 
If the tilt angle $\theta$ satisfies Laplace's equation, the second
term on
the right hand side of (\ref{eq_green_th}) vanishes,
and the free energy becomes
\begin{equation}
F_{el}=\frac{K}{2}\oint_{\partial\Omega}\theta\frac{\partial\theta}{\partial {\bf n}}d{\bf l} \;\;.
\label{eq_contour_en}
\end{equation}

Close to the outer boundary, for $r\gg a$, the director is
nearly uniform and the contribution of the integral over this contour
may be neglected.
The contribution of the integral over the upper disk's 
contour is calculated by taking $\gamma_1(t)=(x(t),y(t))=(a\cos
t,a\sin t)$ with $\theta(t)=t$, yielding 
\begin{eqnarray}
\int_{\gamma_1}\theta\frac{\partial\theta}{\partial {\bf n}}d{\bf l}=
-\pi \Big [\ln \Big (1-\frac{a}{r_d} \Big )+\ln \Big (1+\frac{a}{r_d} \Big) \Big]
\label{eq_gamma_1_en}
\end{eqnarray}
By symmetry the integral over the lower disk's contour is identical.

Now consider the integrals over the straight lines along the $x$-axis
from the disk to the defects.
On $\gamma_2$ and $\gamma_6$, the tilt angle is zero and there is no 
contribution to the free energy.
On $\gamma_4(t)=(t,0)$ (and $\gamma_8$), $\theta(t)=\pi$ and a simple
calculation yields
\begin{eqnarray}
\int_{\gamma_{4}}\theta\frac{\partial\theta}{\partial {\bf n}}d{\bf l}
\approx&2\pi \Big (2\ln\frac{r_d}{a}-\frac{1}{2}\ln\frac{\xi}{r_d-a}-\frac{1}{2}\ln\frac{2r_d}{r_d+a}\nonumber \\
&-\frac{1}{2}\ln\frac{r_d+a}{a}-\frac{1}{2}\ln\frac{r_d^2+a^2}{a(r_d+a)}\Big ) \;\;.
\label{eq_gamma_rl_en}
\end{eqnarray}
in the limit $r_d >> \xi$. If $a >> \xi$ we may neglect the 
contributions from the circular contours
around the defects. Finally, the Frank elastic free energy is given by
\begin{equation}
F_{el}(r_d)\approx\frac{\pi K}{2} \Big[\ln\frac{a}{\xi}-\ln 2 \Big(
\frac{a^3}{r_d^3}-\frac{a^7}{r_d^7} \Big ) \Big]\;\;.
\label{eq_quadrupolar_energy}
\end{equation}
Minimizing with respect to $r_d$ yields $(r_d)_{eq}=\sqrt[4]{7/3}\;a$,
as obtained by Fukuda and Yokoyama \cite{Fukuda} using a simple force 
balance argument. 
The total free energy for this quadrupolar configuration is obtained by 
adding the core energy of the defects $2F_c=\pi K/4$,
and using the appropriate core's size $\xi=\zeta/2$.

The Frank elastic free energy for a dipolar configuration, with a
topological defect of charge $q=-1$, with the preferred position at
${\bf r}_d=(0,r_d)$ was calculated by Pettey et. al. \cite{Pettey}:
\begin{equation}
F_{el}(r_d)\approx\pi K \Big [\ln\frac{a}{\xi}-\ln \Big (\frac{a^2}{r_d^2}-\frac{a^4}{r_d^4} \Big )\Big ]
\label{eq_dipolar_energy}
\end{equation}
yielding the equilibrium position $(r_d)_{eq}=\sqrt{2}\;a$.
The core energy of the defect is now $F_c=\pi K/2$, 
and the core's size is $\xi=\zeta$.

The difference between the free energy of these configurations, 
when the core energies of the defects are taken into account, 
is a function of the correlation length $\zeta$:
\begin{eqnarray}
F_{dip}-F_{quad}
&\approx &\frac{\pi K}{2} \Big (2.08+\ln\frac{a}{\zeta} \Big)\;\;.
\label{eq_difference_energy}
\end{eqnarray}
For $a>\zeta$, the difference is positive, and thus the quadrupolar
configuration has the lowest free energy.

This result contrasts with that for spherical colloidal particles
in 3D nematic liquid crystals.
In the latter case, the dipolar configuration has a punctual defect (with zero core energy).
The quadrupolar configuration exhibits a disclination ring,
and its elastic free energy depends logarithmically on the ratio
between the correlation length and the particle's radius.
For small particles, the nematic distortion is of the quadrupolar type.
As we increase the ratio $a/\zeta$, the energy of the
quadrupolar configuration increases, and eventually exceeds that 
of the dipolar configuration.
This crossover occurs at $a/\zeta\sim 10^3$, that
corresponds to particles with radii of the order of $\mu$m.

\section{Long range interaction}

If the separation ${\bf R}=(R\cos\alpha,R\sin\alpha)$ between particles is 
large, the nematic distortion will be approximately the sum
of the isolated quadrupolar solutions, $\theta\approx\theta_1+\theta_2$,
with an interaction energy 
\begin{eqnarray}
F_{int}&\approx&K\int_\Omega\nabla\theta_1\nabla\theta_2d^2{\bf r}\nonumber\\
&=&K\oint_{\Gamma_1}\theta_1\frac{\partial\theta_2}{\partial {\bf n}}d{\bf l}
+K\oint_{\Gamma_2}\theta_2\frac{\partial\theta_1}{\partial {\bf n}}d{\bf l}
\label{eq_interaction_energy}
\end{eqnarray}  
where the subscripts refer to the different disks (see Fig.2).

\begin{figure}[ht]
\par\columnwidth=20.5pc
\hsize\columnwidth\global\linewidth\columnwidth   
\displaywidth\columnwidth
\centerline{\epsfxsize=180pt\epsfbox{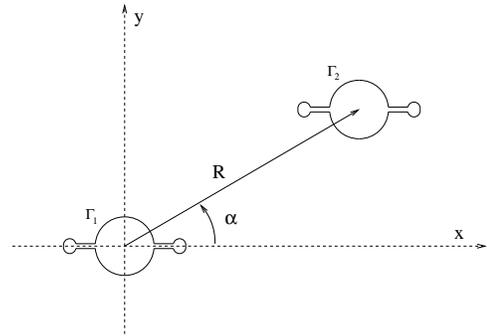}}
\caption
{Two particle configuration: notation and reference frame used in the 
analysis.}
\label{fig_f2}
\end{figure}

Consider the integral over the contour $\Gamma_2$
around the second disk. We use the same notation as for isolated
disks and neglect the contributions from the circular contours around the 
defects.
To calculate the contribution from the upper circular contour around disk 
2 we write $\gamma_{1}(t)=(R\cos\alpha+a\cos t,R\sin\alpha+a\sin t)$.
On this contour $\theta_2(t)=t$, and $\theta_1$ can be expanded as a power
series in $a/R$ yielding
\begin{eqnarray}
\int_{\gamma_{1}}\theta_2\frac{\partial\theta_1}{\partial {\bf n}}d{\bf
l}&=&
\int^0_\pi t \Big (\left.\frac{\partial\theta_1}{\partial x}\right|_{\bf
R}a\cos t
+\left.\frac{\partial\theta_1}{\partial y}\right|_{\bf R}a\sin t \nonumber\\
&+&\left.\frac{\partial^2\theta_1}{\partial x^2}\right|_{\bf R}a^2\cos^2t
+\left.\frac{\partial^2\theta_1}{\partial y^2}\right|_{\bf R}a^2\sin^2t \nonumber\\
&+&2\left.\frac{\partial^2\theta_1}
{\partial x\partial y}\right|_{\bf R}a^2\sin t\cos t \Big )dt+...
\label{eq_gamma1_integral}
\end{eqnarray}  
When the contribution from the lower circular contour around the disk
is added, some of the terms in (\ref{eq_gamma1_integral}) cancel.
After integration, the sum of these two contributions gives
\begin{equation}
\int_{\gamma_{1}+\gamma_{5}}\theta_2\frac{\partial\theta_1}{\partial {\bf
n}}d{\bf l}=
\pi a^2\left.\frac{\partial^2\theta_1}{\partial x\partial y}\right|_{\bf R}+...
\label{eq_gamma15_integral}
\end{equation}  
The integrals over the straight lines can be computed in a similar manner.
On $\gamma_{2}$ and $\gamma_{6}$ the tilt angle $\theta_2=0$, 
and there is no contribution to the free energy.
On $\gamma_{4}(t)=(R\cos\alpha+t,R\sin\alpha)$ (or $\gamma_{8}$), 
$\theta_2(t)=\pi$ and $\theta_1$ may be expanded as a power series in
$a/R$. The result is
\begin{equation}
\int_{\gamma_{rl}}\theta_2\frac{\partial\theta_1}{\partial {\bf n}}d{\bf l}=
\pi (r_d^2-a^2)\left.\frac{\partial^2\theta_1}{\partial x\partial y}\right|_{\bf R}+...
\label{eq_gammarl_integral}
\end{equation} 
The derivative $\partial^2\theta_1/\partial x\partial y$ is obtained from
the quadrupolar expression
\begin{equation}
\theta_1=\frac{\pi}{2}-(r_d^2+\tilde{r}_d^2)\frac{xy}{(x^2+y^2)^2}+...
\label{eq_quadrupolar}
\end{equation} 
valid for large $r=\sqrt{x^2+y^2}$. Finally, adding
the symmetric contribution from the contour $\Gamma_1$, the
total interaction energy becomes
\begin{equation}
F_{int}\approx 6\pi K(r_d^4+a^4)\;\frac{1-2\sin^2 2\alpha}{R^4}\;\;.
\label{eq_interaction_result}
\end{equation} 
The effective interaction between disks decays as $R^{-4}$
and is strongly anisotropic: repulsive if the particles are aligned
horizontally
or vertically ($\alpha=0$ or $\alpha=\pi/2$, respectively), and attractive
for intermediate oblique orientations (the preferred orientation is 
$\alpha=\pi/4$).

\section{Short range interaction}

At small disk separations, the nematic 
deformation is no longer the sum of the isolated quadrupolar solutions 
(\ref{eq_interaction_result}).

The interaction between two circular disks was studied numerically,
using finite elements with adaptive meshes to minimize the total nematic free 
energy at a given separation.
A first triangulation respecting the predefined physical geometry 
(disks separation and orientation) is constructed \cite{BL2D}.
The triangulation space corresponds to 
the domain of integration of the free energy.
It is delimited by the disks' boundaries and an outer boundary at infinity, 
in practice a large distance from the disks, of the order of $20 a$.
Although analytically simpler, the Frank free energy
(\ref{eq_frank_energy}) is plagued with 
divergences of $\nabla\theta$ at the defects,
and thus it is not the most adequate for numerical computations.
For this reason, we used the Landau-de Gennes free energy
(\ref{eq_tensor_energy}) for a 2D nematic. 
The functions $Q_{xx}$ and $Q_{xy}$ are set at all vertices
of the mesh and are linearly interpolated within each triangle.
Using standard numerical procedures the elastic free energy is minimized
under the constraints imposed by the boundary conditions, i.e.,
homeotropic strong anchoring at the disks perimeter and
constant alignment at the outer boundary.
Finally, to overcome the differences in length scales
set by the disks and the defects, of up to 2 orders of magnitude 
(we chose $\zeta^2\sim L/A=10^{-3}$),
new adapted meshes are generated iteratively from the result of the 
previous minimization.
The new local triangle sizes are related with the variations of the
previous solution.
If the variations are strong, finer meshes are required 
in order to guarantee an almost constant numerical weight of each
minimization variable \cite{Patricio}.
The final meshes with $2\times 10^4$ variables and
a minimal edge length of $10^{-4}a$ provided excellent accuracy,
yielding free energies with a relative error of the order of $10^{-4}$.

To test the numerical accuracy of the procedure we calculated the
nematic configuration for a single isolated disk, and found a solution
with two half integer defects at $r_d=1.23a\pm 0.01a$, 
in very good agreement with the analytical result.

\begin{figure}[ht]
\par\columnwidth=20.5pc
\hsize\columnwidth\global\linewidth\columnwidth   
\displaywidth\columnwidth
\centerline{\epsfxsize=190pt\epsfbox{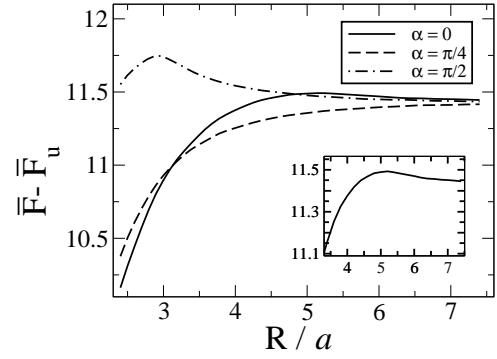}}
\caption
{Reduced elastic free energy $\overline{F}=F/K$ 
as a function of the distance $R$ between the disks,
at three different orientations ($\alpha=0, \pi/4, \pi/2$).
$F_u=F[Q=Q_{eq}]$.}
\label{fig_f3}
\end{figure}

The effective interaction between two disks is plotted in
Fig.3 as a function of the distance $R$ between the disks,
at three different orientations ($\alpha=0, \pi/4, \pi/2$).
A simple fit of these curves confirmed
the quadrupolar long range decay $R^{-4}$, at large disk separations.
However, for smaller separations, the free energy changes dramatically.
In particular, orientations that are repulsive at large separations
($\alpha=0,\pi/2$) become attractive at different threshold distances 
($R_{th}\approx 5a$ and $R_{th}\approx 3a$, respectively). 
Also, the free energy corresponding to the parallel orientation
($\alpha=0$) becomes smaller than the free energy
of the long range preferred oblique orientation ($\alpha=\pi/4$) at short 
distances.

To elucidate this behavior
we plotted on the left of Fig.4 the elastic free energy
as a function of the orientation $\alpha$, 
at three different separations ($R=2.4a,3.0a,4.0a$).
The minimum of each curve defines
a preferred orientation $\alpha^*$, that decreases as the disks approach
each other.
On the right of Fig.4 we see that at very large separations, 
the disks have an oblique orientation (with $\alpha^*=\pi/4$),
that changes to a parallel orientation (with $\alpha^*=0$)
as their separation decreases.

\begin{figure}[ht]
\par\columnwidth=20.5pc
\hsize\columnwidth\global\linewidth\columnwidth   
\displaywidth\columnwidth
\centerline{\epsfxsize=240pt\epsfbox{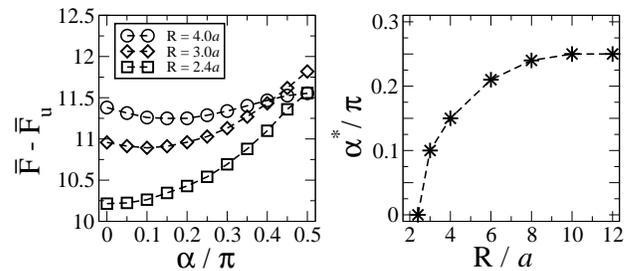}}
\caption
{Left: Reduced elastic free energy
as a function of the orientation $\alpha$, 
at three different separations ($R=2.4a,3.0a,4.0a$). 
Right: Preferred orientation $\alpha^*$ as a function of the distance $R$.}
\label{fig_f4}
\end{figure}

In the following figures we show the nematic order parameter
for several separations, when the orientation of the disks is parallel 
($\alpha=0$, Fig.5)
or perpendicular ($\alpha=\pi/2$, Fig.6).
The nematic order parameter varies between $Q=0$ 
(grey regions) and $Q=Q_{eq}$ (white). These regions
around the disks correspond to the nematic defects.

\begin{figure}[ht]
\par\columnwidth=20.5pc
\hsize\columnwidth\global\linewidth\columnwidth   
\displaywidth\columnwidth
\centerline{\epsfxsize=180pt\epsfbox{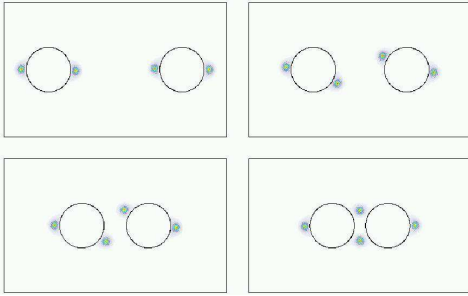}}
\caption
{Nematic configurations for several separations and horizontal alignment 
$\alpha=0$.
The nematic order parameter varies between $Q=0$ 
(grey regions) and $Q=Q_{eq}$ (white).}
\label{fig_f5}
\end{figure}

In both disk alignments, 
if the separation is large enough, the half integer defects around
each disk stay at the same positions as in the isolated disk
configuration.
However, at a certain separation $R_{th}$, exactly when the interaction
becomes attractive for that orientation, the position of the defects
starts to change, in order to minimize the total free energy.

\begin{figure}[ht]
\par\columnwidth=20.5pc
\hsize\columnwidth\global\linewidth\columnwidth   
\displaywidth\columnwidth
\centerline{\epsfxsize=150pt\epsfbox{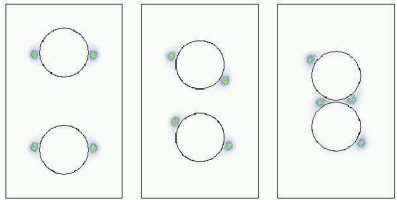}}
\caption
{Nematic configurations for several separations and perpendicular 
alignment $\alpha=\pi/2$.}
\label{fig_f6}
\end{figure}

Finally, 
we analyzed in detail the interaction between the disks for very small
separations.
For strong (fixed) homeotropic anchoring, 
we observed a repulsion when the disks are nearly at contact (for all
orientations) at a separation $R\approx 2.1a$.
This repulsion may not occur if we do not impose
fixed homeotropic boundary conditions at the disks. 
If we consider a finite anchoring strength we have to add to the 
free energy (\ref{eq_tensor_energy}) a surface term
\begin{equation}
F_a=\frac{W}{2}\int_{\Lambda_1+\Lambda_2}(Q_{ij}-Q_{ij}^n)^2 dl
\end{equation}
where $Q_{ij}^n$ are
the values of the tensor order parameter for homeotropic alignement.

In Fig.7 we plot the total elastic free energy 
as a function of the distance $R$ between disks, 
for three different reduced anchoring strengths 
($W/Ka=1000; 40; 30$).
These anchoring strengths are large enough to induce defects
in the nematic host.
Clearly the repulsion appears at small separations for
strong anchoring ($W/Ka=1000$) and vanishes at a critical anchoring
strength that  
lies between $30<W/Ka<40$. 
In the weak anchoring regime the coalescence of the dropplets may occur.

\begin{figure}[ht]
\par\columnwidth=20.5pc
\hsize\columnwidth\global\linewidth\columnwidth   
\displaywidth\columnwidth
\centerline{\epsfxsize=230pt\epsfbox{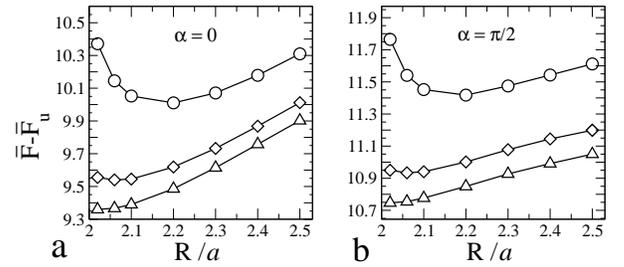}}
\caption
{{Reduced elastic free energy as a function of the disks separation $R$,
for three reduced anchoring strengths ($W/Ka=1000; 40; 30$), corresponding 
to the circles, diamonds and triangles respectively}. Left: Parallel
alignment ($\alpha=0$).
Right: Perpendicular alignment ($\alpha=\pi/2$).}
\label{fig_f7}
\end{figure}
 
\section{Conclusions}

We compared the (analytic) elastic free energies of 2D nematic
configurations with one (satellite) and two (ring) defects around isolated 
disks with homeotropic boundary conditions, and found that the ring
configuration is the stable one for realistic values
of the reduced correlation length $\zeta/a$.

Using the analytic ring solution we derived
the long range interaction between two disks immersed in
2D nematic hosts. We found an anisotropic interaction,
that is repulsive for orientations close to perpendicular or parallel
orientations and attractive for oblique alignement, with a preferred
orientation at $\alpha=\pi/4$.
In this limit, the interaction decays as $R^{-4}$.

Finally, the analytic results were compared with very
accurate numerical calculations, valid at arbitrary separation.
At small disk separations, the defects may change their positions,
leading to an unexpected complex interaction with the long range repulsive
orientations becoming attractive.
As a result as we decrease the distance between the disks,
their preferred orientation changes from oblique to parallel 
alignment.
For sufficiently large anchoring strengths, at very short range, we
found a new repulsion for all orientations, that may prevent coalescence.
This repulsion vanishes for small anchoring strengths.

The equilibrium structure of inverted nematic emulsions is 
complex and depends sensitively on the radii of the disks,
the nematic correlation length and anchoring strength. In particular  
short range nonlinear effects may change the interaction between 
disks dramatically. It is thus reasonable to assume that three
body interactions may be relevant to describe the formation of the
the structures observed experimentally.
Finally, in order to understand the observed structures  
dynamical aspects may also have to be taken into account.

{\bf Acknowledgments}

We acknowledge the support of the Funda\c c\~ao para a 
Ci\^encia e Tecnologia (FCT) through a running grant
(Programa Plurianual) and grants 
No. SFRH/BPD/1599/2000 (MT) and No. SFRH/BPD/5664/2001 (PP).

\end{multicols}

\end{document}